\documentstyle[12pt]{article}
\input{epsf.tex}
\def\lsim{\mathrel{\rlap {\raise.5ex\hbox{$<$}}
{\lower.5ex\hbox{$sim$}}}}

\begin{document}
\begin{titlepage}
\begin{flushright}
CERN-TH/96-96\ \\
hep-ph/xxxxxxx{\hskip.5cm}\\
\end{flushright}
\begin{centering}
\vspace{.3in}
{\bf GUTs WITH EXCLUSIVELY $\bf \Delta B=1$,
 $\bf \Delta L=0$ \space   $\bf R$-PARITY VIOLATION} \\
\vspace{2 cm}
{K. TAMVAKIS$^{\ast}$}
\vskip .5cm
{\it Theory Division, CERN}\\
{\it 1211 Geneva 23, Switzerland}\\
\vspace{1.5cm}
{\bf Abstract}\\
\end{centering}
\vspace{.1in}
We study $R$-parity violation in the framework of GUTs,
 focusing on the 
case that $R$-parity is broken exclusively through
 $\Delta B=1$, $\Delta L=0$ effective interactions.
We construct two such models, an $SU(5)$ and 
an $SU(5)\times U(1)_{X}$ model, in 
which $R$-parity breaking is induced 
through interactions with extra supermassive fields.
 The presence of only the Baryon Number 
violating operators ${d^c}{d^c}{u^c}$
 requires an asymmetry between quarks and leptons, 
which is achieved either by virtue of the Higgs 
representations used or by modifications in the 
matter 
multiplets. The latter possibility is realized in the 
second of the above models, where the left-handed 
leptons 
have been removed from the representation in which they 
normaly 
cohabit with the right-handed up quarks and enter as a 
combination of the isodoublets in
 $(\bf {{\overline{5}},-3})$
 and 
$(\bf {5,-2})$ representations. 
In both models the particle content 
below the GUT scale is unaffected by the
 introduced $R$-parity 
breaking sector.
\vspace{2cm}
\begin{flushleft} CERN-TH/96-96\\
April 1996\\
\end{flushleft}
\hrule width 6.7cm \vskip.1mm{\small \small $^\ast$\ On leave from 
Physics Department, University of Ioannina, GR-45110, Greece.}
\end{titlepage}
\newpage

\begin{bf}1. Introduction.\end{bf} In contrast to the
 Standard Electroweak Model where $B$- and $L$-number 
conservation is automatic for the minimal field content,
 in the Supersymmetric Standard Model\cite{NHK}
 renormalizable interactions among the standard chiral
 matter superfields 
can be present\cite{WSY} which violate both $B$ and $L$.
 These interactions are 
\begin{equation}{\lambda_{ijk}l_il_je^c_k + \lambda'_{ijk}d^c_il_jq_k + \lambda''_{ijk}d^c_id^c_ju^c_k + \epsilon_il_iH \,.}\end{equation}
The indices are generation indices. These terms can
 be avoided by imposing a discrete 
symmetry called $R$-parity and defined as
 $R=(-1)^{3B+L+2S}$, $S$ being the spin.
If $R$-parity is not an exact symmetry and the
 above interactions are present\cite{ZHR}, the 
combination 
of the second and the third term in (1) results 
in proton 
decay through squark exchange at an unacceptable rate,
 unless the product of these 
couplings is extraordinarily small, i.e.
 $\lambda'\lambda''\leq 10^{-24}$.
If one is restricted to the Supersymmetric
 Standard Model, the above four couplings are 
independent, and it would be technically possible
 to assume the existence of some of them while 
putting to zero others or setting them to very small values.
This is something that cannot be done in GUTs, at least
 in a straightforward fashion \cite{SV}.
For example, in minimal $SU(5)$ the first three terms in
 (1) arise from the $R$-parity violating 
 coupling \begin{equation}{\lambda_{ijk}\phi_i(\overline{5})\phi_j(\overline{5})\psi_k(10) \,.}\end{equation}
Thus, the resulting couplings in (1) would be
 related as $\lambda_{ijk}=\frac{1}{2}\lambda'_{ijk}=\lambda''_{ijk}$.
It is possible however that $R$-parity violation
 is absent at the renormalizable level, perhaps
 because of some 
other symmetry \cite{KT} not directly related to it,
 and shows up in the form of effective 
non-renormalizable interactions suppressed by 
the breaking scale of the symmetry over some large
 mass scale. The required smallness of these 
couplings, coming 
mainly from the need to suppress proton decay, 
can be accounted for by establishing a large hierarchy 
among the $B$- and $L$-violating 
effective strengths. $SU(5)$ models with the requred
 $\lambda''\ll\lambda'$ disparity in the effective
 $R$-parity non-conserving 
strengths have been recently discussed \cite{SV},\cite{KT}.
 In contrast to $R$-parity non-conservation \cite{DRG} 
through $L$-number 
violation $(\lambda''\ll\lambda')$, which 
has received considerable attention, 
the opposite case of $R$-non-conservation exclusively 
through the Baryon Number violating interaction \begin{equation}{\lambda''_{ijk}d^c_id^c_ju^c_k}\end{equation}
has received much less attention and only within the 
Supersymmetric 
Standard Model. The phenomenological profile of low-energy 
Baryon Number 
violation through (3) includes neutron--antineutron 
oscillations, 
double nucleon decay (in nuclei), as well as 
various exotic
 non-leptonic 
heavy meson decays.
The presence of (3), while all the other terms in (1)
 are absent, clearly requires 
a strong asymmetry between quarks and leptons, which cannot 
be accounted for in conventional 
GUTs. Nevertheless, such GUTs can certainly be constructed.
In the present short article we construct two such models, an 
$SU(5)$ model and a flipped $SU(5)\times U(1)_X$ model,
 in which $R$-parity 
is violated exclusively through the operators (3).
 These models, although entirely realistic, 
are only technically natural, as all existing GUTs. They 
demonstrate that 
gauge coupling unification, an almost ``experimental" fact, 
is certainly compatible with an extreme disparity between 
quarks and leptons as far as their $R$-parity violation 
behaviour is concerned.

\begin{bf}2. An $\bf {SU(5)}$ model.\end{bf} The so-called
 {\it missing-doublet $SU(5)$ model} \cite{MG} was 
constructed 
in order to avoid the fine numerical adjustment in the 
triplet--doublet mass 
splitting required in the minimal supersymmetric $SU(5)$ 
model \cite{SG}. The 
generic missing-doublet $SU(5)$ model has a Higgs 
superfield $\Sigma $ in the 
{\bf 75} representation, instead of the usual 
adjoint and an extra pair of 
superfields $\Theta , \overline{\Theta } $ in 
the ${\bf 50}+{\bf {\overline{50}}}$ representation.
The superpotential 
is \begin{equation}W=Y^{(u)}_{ij}\psi_i \psi_j H +
Y^{(d)}_{ij}\psi_i \phi_j \overline{H} +
 \lambda \overline{H} \Sigma \Theta +
 \overline{\lambda} H \Sigma \overline{\Theta} 
+\frac{1}{2} \mu_{\Sigma}  Tr(\Sigma^2) + 
\frac{h}{3}Tr(\Sigma^3) \end{equation}
where $\psi_i(10)$, $\phi_i(\overline{5})$ 
are the standard three families and 
$H(5)=(H^c, D)$, $\overline{H}(\overline{5})=(H, D^c)$
 the electroweak Higgs pentaplets.
 Note that the $SU(5)$-breaking v.e.v. 
$\langle \Sigma \rangle $ pairs the 
coloured triplets $D$, $D^c$ to 
the analogous coloured triplets $\theta$, $\theta^c$ in 
the $\bf 50$+$\bf \overline{50} $.
The rest of the ingredients of the
 $\bf 50$+$\bf \overline{50} $ can 
receive a mass 
either through a direct term 
$M_{\Theta} \Theta \overline{\Theta} $ or through 
a coupling $\Theta \Sigma \overline{\Theta} $.
 In the latter case, the gauge 
coupling blows up before we reach the Planck mass 
$M_P$ and the model ceases to 
be perturbative. The same happens if $M_{\Theta}$ 
is of the order of the unification 
scale. If $M_{\Theta}$ is of the order of $M_P$, 
perturbativity is valid. In this 
case one pair of triplets, via a see-saw type 
mechanism, receives a mass of order 
${\langle\Sigma\rangle }^2/M_{\Theta}$ ,
 which is two orders of magnitude 
below the unification scale and, therefore,
 problematic for proton stability due 
to the presense of $D=5$ operators. 
This problem is avoided in the Peccei--Quinn 
version of the model\cite{HMT} which 
starts with two pairs of pentaplets and two 
pairs of Planck-mass 
$\bf 50$+$\bf \overline{50}$'s and 
ultimately ends up with 
an additional pair of intermediate mass
 ($10^{10}$-$10^{12}$ GeV) isodoublets.
What we are about to discuss in relation 
to $R$-parity applies equally well to 
either version of the model. Thus, for simplicity
, we shall be referring to the 
generic superpotential (4). Note however that 
the Peccei--Quinn version is in 
agreement with low energy data \cite{HMT}.

 The superpotential (4) is exactly $R$-parity 
conserving. Let us introduce now an 
extra sector of  massive fields $R(50)+
\overline{R}(\overline{50})+
\eta(5)+\overline{\eta}(\overline{5})$
in more than one family replicas. All these 
fields will have $O(M_P)$ masses so 
that perturbativity and particle content below 
$M_P$ will
 be unaffected. The new sector breaks $R$-parity
 through the interactions \begin{equation}\Delta W_R=
\lambda_i\phi_i\Sigma R_i 
+f_{ijk}\psi_i\overline{\eta}_j\overline{\eta}_k + 
f_i\eta_i\Sigma\overline{R}_i +M_{Ri}R_i \overline{R}_i +
M_{{\eta}i}\eta_i \overline{\eta}_i \,.\end{equation}
It is evident that for a SM-preserving v.e.v. 
of $\Sigma$ , the left-handed leptons 
in $\phi$ will not communicate with the contents 
of $\psi$ through the interactions (5), since 
$R$ contains only coloured components and a charged 
isosinglet.

 Denoting by $\Delta_{0i}$, $\Delta^{c}_{0i}$ 
the triplets in $R_i$, $\overline{R}_i$ and by
 $\delta_{0i}$, $\delta^{c}_{0i}$ the 
corresponding ones 
in $\eta_i$, $\overline{\eta}_i$ we obtain 
the triplet mass matrix 
\begin{equation}M^{(3)}=\left[\begin{array}{rrr}
M_R&fv&0\\
0&M_{\eta}&0\\
\lambda v&0&0
\end{array}\right]\end{equation}
in a $\Delta^{c}_0$, $\delta^{c}_0$, 
$d^{c}_0$/$\Delta_0$, $\delta_0$
 basis. The combination 
\begin{equation}d^c=N(d^{c}_{0}-(\lambda v/M_R)\Delta^{c}_{0} +
(\lambda f v^{2}/M_R M_{\eta})\delta^{c}_{0})\end{equation}
with $N^{-1/2}=1+(\lambda v/M_R)^{2}(1+(fv/M_{\eta})^{2})$, stays 
massless. Rewriting (5) in terms of the 
mass eigenstates and integrating out 
the massive ones, we obtain an effective 
non-renormalizable interaction 
term \begin{equation}f^{eff}_{ijk}(v/M)^{4}(u^{c}_id^{c}_jd^{c}_k)\end{equation}
which violates $R$-parity exclusively through
 the quark superfields.

 Another way to understand the effective 
interaction (6) is through the 
 graph of the figure, which generates the
 effective F-term 
\begin{equation}(\psi_k)_{MN}(\phi_i)^{M^\prime}
(\phi_j)^{N^\prime}\omega^{MN}_{M^\prime N^\prime}\end{equation}
with $\omega^{MN}_{M^\prime N^\prime}=
\omega^{M}_{M^\prime} \omega^{N}_{N^\prime}$ and 
\begin{equation}\omega^{A}_{B}=
\epsilon^{ACDEF} \epsilon_{BGHIJ} \Sigma^{GH}_{LM} \Sigma^{PQ}_{CD} (\delta^{L}_{P} \delta^{I}_{E} \delta^{M}_{Q} \delta^{J}_{F} 
+\cdots )\,.\end{equation}
The dots imply symmetrization with respect to 
$L$, $I$ and $M$, $J$ and 
antisymmetrization with respect to $L$, $M$ and $I$, 
$J$. Substituting the 
$SU(3)_{C}\times SU(2)_{L} \times U(1)_{Y}$ invariant v.e.v.
\begin{equation}\langle\Sigma^{BC}_{DE}\rangle
=v((\delta_{c})^{B}_{D}(\delta_{c})^{C}_{E} + 2(\delta_{w})^{B}_{D}(\delta_{w})^{C}_{E}-
\frac{1}{2}\delta^{B}_{D}\delta^{C}_{E} -(B\leftrightarrow C))\end{equation}
we end up with \begin{equation}\omega^{A}_{B}\propto v^{2}(\delta_{c})^{A}_{B}\,.\end{equation}
The subscript {\it c } denotes the $SU(3)_{C}$ direction.

 Note that (5) is only technically natural. 
Even in the Peccei--Quinn 
version of 
the model, the symmetries
 allow $R$, $\overline{R}$ to be replaced by $\Theta$,
 $\overline{\Theta}$. This should not be 
allowed, however, and the $R$-parity 
sector should not interact with the rest of 
the theory apart from the 
interactions appearing in (5).

\begin{centering}
\begin{figure}
\leavevmode
\epsfbox{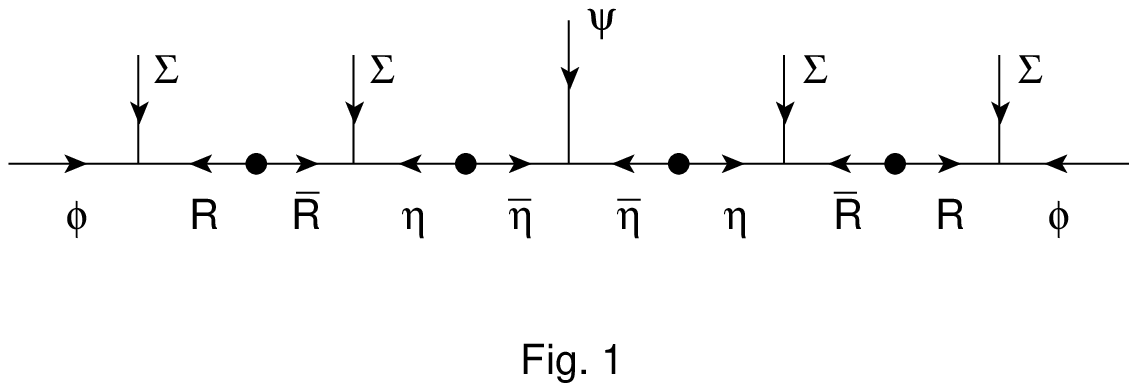}
\end{figure}
\end{centering}

\begin{bf}3. An $\bf {SU(5)\times U(1)_{X}}$ model.\end{bf} 
In the previously analysed $SU(5)$ model, 
the $R$-parity non-conserving interactions 
were restricted to quark superfields by virtue
 of the choice of the 
Higgs representations employed. Another possibility
 would be to construct GUTs 
with a built-in quark--lepton asymmetry by virtue 
of modifications
 in the matter 
representations themselves. Such a 
{\it de}-{\it unification} is already partially 
realized in the flipped $SU(5)\times U(1)_{X}$ 
model \cite{BDK}, in which the 
right-handed leptons have been removed from the 
rest of the quark and lepton 
representations and are introduced as $SU(5)$ singlets
. As a result, no relation 
between quark and charged lepton masses exists in 
this model. In what follows
we shall construct an $SU(5)\times U(1)_{X}$ model 
with exclusively $\Delta B=1$,
 $\Delta L=0$ $R$-parity violating interactions. 
Our strategy will be to remove 
the left-handed leptons from the $\bf (\overline{5},-3)$ 
representation in which 
they cohabit with the up antiquarks. Then, $R$-parity
 non-conserving 
interactions of the type $d^{c}d^{c}u^{c}$ will not 
necessarilly coexist with 
those of the $d^{c}ql$ type.

 The standard Higgs fields of the flipped 
$SU(5)\times U(1)_{X}$ model
are $$ {\cal H }({\bf 10,1}) 
+\overline{\cal H}({\bf \overline{10},-1}) 
+ h({\bf 5,-2}) 
+ \overline{h}({\bf\overline{5},2}) = $$
\begin {equation}(q_{H}, d^{c}_{H}, \nu^{c}_{H}) + 
(\overline{q}_{H}, \overline{d}^{c}_{H}, \overline{\nu}^{c}_{H}) 
+(H, D) 
+ (H^{c}, D^{c})\,.\end{equation}
The breaking 
$SU(5)\times U(1)_{X}\rightarrow SU(3)_{C}
\times SU(2)_{L}\times U(1)_{Y}$ is achieved 
with a v.e.v. in the $D$-flat 
direction $\langle\nu^{c}_{H}\rangle=
\langle\overline{\nu^{c}_{H}}\rangle=v\,.$
The coloured triplets $d^{c}_{H}$, $\overline{d^{c}_{H}}$ 
that survive the 
Higgs phenomena combine into massive states with 
the triplets $D$, $D^{c}$ 
through the couplings 
\begin{equation} W_{1}=\lambda_{1}{\cal H}{\cal H} h +
\lambda_{2}{\overline{\cal H} }
{\overline{\cal H}}{\overline{h}} =
(\lambda_{1} v)d^{c}_{H}D 
+(\lambda_{2} v)\overline{d}^{c}_{H} D^{c} 
+\cdots \end{equation}

 Consider now a complete matter family 
\begin{equation} {\cal F}({\bf 10,1})
 +{f^{c}}^{\prime}({\bf {\overline{5}},-3}) 
+e^{c}({\bf 1,5}) =(q, d^{c}, {\nu}^{c})+
(l^{\prime}, {u^{c}}^{\prime})+e^{c}\,.\end{equation}
The prime on ${f^{c}}^{\prime}
({\bf {\overline{5}},-3})$ indicates
 that its contents should not  be identified 
yet with quarks or leptons.
Next, we introduce additional ``matter'' superfields 
\begin{equation}{f^{c}}^{\prime \prime}
({\bf {\overline{5}},-3}) +
f^{\prime \prime}({\bf {5,3}}) +
{\phi}^{c}({\bf {\overline{5},2}})+\phi ({\bf {5,-2}}) =
(l^{\prime \prime}, {u^{c}}^{\prime \prime})+
({l^{c}}^{\prime \prime}, u^{\prime \prime})+
({\lambda}^{c}, {\delta}^{c})+
({\lambda}, {\delta})\,.\end{equation}
The component fields transform under 
$SU(3)_{C}\times SU(2)_{L}\times U(1)_{Y}$ as
 $(\bf{1,2,-\frac{1}{2}})$ and 
$(\bf{1,2,\frac{1}{2}})$ isodoublets,
 namely $(l^{\prime \prime},{\lambda})$ and
 $({l^{c}}^{\prime \prime},{\lambda}^{c})$ 
correspondingly, and as coloured  triplets 
\begin{equation}{u^{c}}^{\prime \prime}({\bf {{\overline{3}},1,-\frac{2}{3}}}),\,
u^{\prime \prime}({\bf{3,1,\frac{2}{3}}}), \,{\delta}^{c}({\bf{{\overline{3}},1,\frac{1}{3}}}), \,{\delta}({\bf{3,1,-\frac{1}{3}}})\,.\end{equation}
Out of all the fields ${f^{c}}^{\prime},
 {f^{c}}^{\prime \prime}, f^{\prime \prime},
 {\phi}^{c}, \phi$ , only one isodoublet 
and only 
one colour triplet 
of charge $-2/3$ will survive massless. 
In order to achieve this, 
an extra 
massive pair of decaplets ${\cal {H}}^{\prime}(\bf{10,1}) 
+{\overline{\cal H}}^{\prime}
({\bf{{\overline{10}},-1}})$
 has to be introduced. Note that all 
introduced extra fields belong 
to representations that arise in the 
superstring version of the model \cite{AEH}.
 A  four-dimensional superstring model having 
the above field content, as well as the 
interactions of the GUT at hand, could 
in principle be constructed.

 These fields interact through the 
superpotential interactions 
$$W_{2}=M_{1}{\overline{\cal H}}^{\prime} 
{\cal H}^{\prime} +
M_{2}f^{\prime \prime}{f^{c}}^{\prime \prime} 
+{\lambda}_{3}{\cal H}^{\prime}{\cal H}{\phi} 
+{\lambda}_{4}{\overline{\cal H}}^{\prime}
{\overline{\cal H}}^{\prime}{\phi}^{c}
+{\lambda}_{5}{f^{c}}^{\prime}{\phi}^{c}{\cal H} 
+{\lambda}_{6}f^{\prime \prime} 
{\phi}{\overline{\cal H}} =$$
$$M_{1}({\overline{q}^{\prime}_H} {q^{\prime}_H} 
+{{\overline{\nu}}^{c \prime}_H} {\nu^{c \prime}_H}
 +{{\overline{d}}^{c \prime}_H}{d^{c \prime}_H})
 +M_{2}({u^{c}}^{\prime \prime}u^{\prime \prime}
 +{l^{c}}^{\prime \prime}l^{\prime \prime}) $$
\begin{equation}+({\lambda_{3}}v){d^{c \prime}_H}{\delta} +({\lambda_{4}}v){{\overline{d}}^{c \prime}_H}
{\delta}^{c} +
({\lambda}_{5}v){l^{\prime}}{\lambda}^{c} 
+({\lambda}_{6}v){l^{c}}^{\prime \prime}{\lambda}\,.\end{equation}
 According to (18) the pairs $({q}^{\prime}_{H}, 
{\overline{q}}^{\prime}_H)$ 
and $(\nu^{c \prime}_H, 
{\overline{\nu}}^{c \prime}_H)$ get a 
mass $M_{1}$ and the pair 
$(u^{\prime \prime}, {u^{c}}^{\prime \prime})$ 
gets a mass $M_{2}$. The 
combinations \begin{equation}(d^{c}_H)_{\pm}=
[({\lambda_{4}}v)d^{c \prime}_H -
M_{\pm}{\delta^{c}}]/{\sqrt{M^{2}_{\pm} + 
({\lambda_{4}}v)^{2}}}\end{equation}
\begin{equation}({\overline{d}}^{c}_H)_{\pm}=
[({\lambda_{3}}v){\overline{d}}^{c \prime}_H -M_{\pm}{\delta}]/{\sqrt{M^{2}_{\pm} +
({\lambda_{3}}v)^{2}}}\end{equation}
get a mass $M_{\pm}={\frac{1}{2}}(M_{1}\pm \sqrt{M^{2}_{1}+4{\lambda_3}{\lambda_4}{v^2}})$. The
 pair $l^{\prime},{\lambda}^{c}$ gets 
a mass ${\lambda_5}v$, so that $f^{c \prime}$ 
does not contain 
any leptons. The combination 
\begin{equation}[M_{2}l^{\prime \prime}
 +({\lambda_{6}}v){\lambda}]
/{\sqrt{M^{2}_{2} 
+({\lambda_6}v)^{2}}}\end{equation}
forms a massive state of mass 
$[M^2_2 +({\lambda_6}v)^2]^{\frac{1}{2}}$
with $l^{c \prime \prime}$. The surviving 
massless 
left-handed lepton is the combination 
\begin{equation}l=[({\lambda_6}v)l^{\prime \prime}
-M_{2}{\lambda}]/{\sqrt{M^2_2 +({\lambda_6}v)^2}}\,.\end{equation}
 Finally, the field $u^{c \prime}$ 
stays massless. Thus, it can be 
identified with an up antiquark and we can 
drop the prime when 
referring to it. It should be admitted that 
the choice of $W_2$ is 
only technically natural and interaction terms 
that would drastically 
change the obtained mass pattern are possible.
 But naturalness is a 
general problem of GUTs.

 It should be pointed out that the key ingredient 
of the present model is 
that ``matter-like" isodoublets and ``Higgs-like" 
ones can obtain superheavy masses through 
the couplings
 ${f^c}{\phi^c}{\cal H}$ . Note that the 
coloured triplets 
contained in the pentaplets are of 
different charge and stay massless. This 
is in a way the 
opposite of 
what happens through the couplings ${\cal H}{\cal H}h $ . 
There, when the decaplet gets a v.e.v., the coloured 
triplets in $\cal H$ 
and $h$ pair to obtain a mass while the isodoublet 
in $h$ is left massless.

 Below the $SU(5)\times U(1)_{X}$ breaking 
scale the model has the 
MSSM particle content. Quark and lepton 
masses arise through the 
standard Yukawa couplings 
\begin{equation}W_{3}=Y^{(d)}{\cal F}{\cal F}h + 
Y^{(u)}{\cal F}{f^{c \prime}}{\overline{h}}
+Y^{(u) \prime}{\cal F}{f^{c \prime \prime}}{\overline{h}}
+Y^{(e) \prime}{f^{c \prime}}{e^c}h 
+Y^{(e)}{f^{c \prime \prime}}{e^c}h\end{equation}
\begin{equation}=Y^{(d)}qHd^c +Y^{(u)}q{u^c}H^c +{\frac{({\lambda_6}v)}{\sqrt{M^2_2+({\lambda_6}v)^2}}}(Y^{(u) \prime}l{\nu^c}H^c +Y^{(e)}lHe^c) +
 \cdots \end{equation}
The dots signify terms that 
involve superheavy fields. The 
coupling $Y^{(e) \prime}$ does not 
contribute to quark--lepton 
masses while $Y^{(u) \prime}$ contributes
 only to a Dirac neutrino 
mass. Note that right-handed neutrinos can 
obtain a large Majorana mass 
through the non-renormalizable interactions
 ${\cal F}{\cal F}{\overline{\cal H}}{\overline{\cal H}}/M=(v^{2}/M){\nu^c}{\nu^c}+\cdots$.
 The mass scales $M_1$ and $M_2$, since they 
are not related to the 
$SU(5)\times U(1)_{X}$ breakdown scale, are not
 necessarily of that 
order. In fact, their natural values are of 
the order
 of the Planck scale or the string scale. In that
 case, lepton masses 
carry a suppression factor ${\lambda_6}v/M_2$.
 Note that in this case the 
triplets $(d^{c}_H)_{-},({\overline{d}}^{c}_H)_{-}$ 
have a mass 
$({\lambda_3}v)^{2}/M_1$, somewhat lower than the 
$SU(5)\times U(1)_{X}$ 
breaking scale. This would only have a very minor
 consequence on the Renormalization Group analysis 
and no other effect since these coloured triplets 
do not appear in the Yukawa couplings 
of quark--lepton bilinears. Of course,
 alternatively
 it is technically
 natural to take the scale 
$M_2$ to be of the same order as ${\lambda}v$.

 Up to now we have considered only one family. 
A three-generation model 
with all three left-handed leptons removed from
 the $(\bf {{\overline{5}},-3})$ 
representations that contain the up antiquarks
 would 
require a triplication of 
the additional sector that has been introduced. 
In the case 
when the scales $M_1$ and $M_2$ are 
of order $M_P$,
 the one-family 
model aquires just one extra pair of 
pentaplets, 
massless above the 
 $SU(5)\times U(1)$ breaking scale. This does not 
have any drastic 
influence on a possibly anticipated unification of 
the $SU(5)$ and 
$U(1)_{X}$ couplings. In the extreme case of
 three extra 
pairs of 
pentaplets above the GUT scale, the $SU(5)$ beta 
function at one loop 
vanishes. Of course, it is possible that the 
left-handed 
lepton ``misplacement" 
occurs only for one generation, possibly the third,
 and that 
the previously 
described sector of massive fields is sufficient.

 $R$-parity is still a symmetry of the 
effective theory 
below the GUT 
scale. Effective operators that could break
 $R$-parity are 
\begin{equation}{\cal F}{\cal F}{\phi}  ,\, 
{\cal F}{\cal F}{\cal H}f^c ,\, 
{\cal H}{f^c}{f^c}e^c \,.\end{equation}
 These operators cannot be generated as 
effective F-terms by the 
interactions appearing in $W_1$, $W_2$ 
and $W_3$. Nevertheless, it is 
straightforward now to introduce 
$R$-parity violation in the desired 
baryonic direction by modifying the 
model so that it contains an extra 
supermassive pair 
of pentaplets \begin{equation}\chi(\bf {5,-2}) 
+ {\chi^c}(\bf {{\overline{5}},2})\end{equation}
interacting with the rest of the theory 
exclusively through the interactions 
\begin{equation}W_{4}=M_{3}{\chi}{\chi^c} 
+{\lambda_{ij}}{{\cal F}_{i}}{{\cal F}_{j}}{\chi} 
+ {\lambda}{\cal H}{f^{c \prime}_k}{\chi^c}\,.\end{equation}
 An effective F-term that involves only quark 
superfields and violates 
$R$-parity can now be generated. It is 
\begin{equation} {\frac{{\lambda}{\lambda_{ij}}}{M_3}}
{{\cal F}_i}{{\cal F}_j}{f^{c \prime}_k}{\cal H}\,.\end{equation}
The index $k$ refers to the generation with 
the misplaced lepton. For example, 
in the case that $k$ corresponds to the third generation, 
the generated effective operator  will
 be ${d^c_i}{d^c_j}{t^c}$ . If we 
assume that no other $R$-parity 
non-conserving interactions are present 
 apart from those appearing in $W_{4}$, 
no other effective F-terms, such as 
${\cal F}{\cal F}{f^{c \prime \prime}}{\cal H}$ 
or ${\cal H}{f^{c \prime}}
{f^{c \prime \prime}}e^c$ , will appear.

\begin{bf}4. Discussion.\end{bf} 
The Baryon Number 
violating operators under discussion 
have various,
in principle testable, phenomenological implications, 
and each of them can provide 
us with information 
on the effective coupling constants involved.
 The effect of these interactions in hadron 
collider experiments is expected to be difficult 
to test since these interactions lead to multijet
 production which suffers from a tremendous 
QCD background.
 Cascade decays however, could lead to more easily 
identifiable signals \cite{DRG}.
Nevertheless, collider processes can be used in 
order to derive bounds on 
these couplings. Considering the contribution of 
these couplings  
to the $Z$ decay into $b,\overline{b}$ or leptons, with 
the present 
state of the data, does not lead to any interesting
 bound \cite{BCS}.
 Both models presented here satisfy trivially these
 bounds. There is 
virtually no cosmological bound on these couplings 
either. Such bounds 
are in general derived by requiring the survival of
 early baryogenesis 
until the present epoch. It has been shown that for
 the exclusively Baryon 
Number violating operators ${d^c}{d^c}{u^c}$ no bound
 is derived and all  
that is required is an initial flavour asymmetric 
Lepton Number asymmetry \cite{DR}. The strongest 
constraints on these couplings come from 
neutron--antineutron 
oscillations and heavy nuclei decays \cite{GS},\cite{BM}.
 Neutron--antineutron 
oscillations constrain the ${d^c}{b^c}{u^c}$ 
coupling while the ${d^c}{s^c}{u^c}$
coupling 
is strongly bounded by the non-observation of 
double-nucleon decay into 
kaons. For squark masses of the order of 300 GeV
, these bounds are 
${\lambda^{\prime \prime}_{udb}}\leq 5\times {10^{-3}}$,
 ${\lambda^{\prime \prime}_{uds}}\leq{10^{-6}}$.
 Additional bounds on products of these couplings 
have been recently \cite{CRS} 
obtained from the consideration of rare two-body 
non-leptonic decays of heavy 
quark mesons (mostly {\it B }).

 The approximate $R$-parity conservation required by
 any phenomenologically 
viable version of the Supersymmetric Standard Model 
is one of the intriguing 
questions of supersymmetric model building. This is
 dramatically encountered 
in Superstring derived models where in
 general $R$-parity
 is not a symmetry 
and special care has to be taken so that it is not
 badly broken. Assuming of 
course that low-energy supersymmetry is realized 
in nature, it might very 
well be that $R$-parity is an exact symmetry.
 Neither Superstrings nor GUTs 
have yet provided any convincing argument 
why it should be so. Thus the 
possibility of $R$-parity non-conservation 
remains open. The models discussed 
in the present article are realistic examples
 of GUTs, i.e.theories realizing 
the gauge coupling unification suggested by 
low-energy electroweak data, 
which at the same time exhibit $R$-parity 
non-conservation. These models 
demonstrate the compatibility of unification
 and $R$-parity breaking 
exclusively through Baryon Number violation.
 Although this type of $R$-parity 
breaking would not necessarily be the easiest 
to observe, its 
rare phenomenological profile  would certainly
 provide evidence for supersymmetry.

\end{document}